\documentclass[twocolumn,aps,pra,showpacs,floatfix]{revtex4-1}
\usepackage{graphicx} 
\usepackage{amsmath}
\usepackage{amssymb}
\usepackage{bm}
\usepackage{color}
\usepackage{hyperref}
\usepackage{dcolumn}

\begin{document}

\title{Lattice-induced non-adiabatic frequency shifts in optical lattice clocks}

\author{K. Beloy}
\affiliation{
Centre for Theoretical Chemistry and Physics,
The New Zealand Institute for Advanced Study,
Massey University Auckland, Private Bag 102904, 0745, Auckland, New Zealand
}

\begin{abstract}
We consider the frequency shift in optical lattice clocks which arises from the coupling of the electronic motion to the atomic motion within the lattice. For the simplest of 3-D lattice geometries this coupling is shown to only affect clocks based on blue-detuned lattices. We have estimated the size of this shift for the prospective strontium lattice clock operating at the 390 nm blue-detuned magic wavelength. The resulting fractional frequency shift is found to be on the order of $10^{-18}$ and is largely overshadowed by the electric quadrupole shift. For lattice clocks based on more complex geometries or other atomic systems, this shift could potentially be a limiting factor in clock accuracy.
\end{abstract}

\date{\today}


\maketitle

\newcommand{\cmt}[1]{[\![#1]\!]}


In 2003, Katori \emph{et al.}~\cite{KatTakPal03} first proposed a frequency standard based on neutral atoms confined in a well engineered optical lattice, citing the potential to combine the benefits of the state-of-the-art clocks based on single-ions in a Paul trap (Doppler-free and recoil-free spectroscopy, long interrogation times) and ensembles of neutral atoms in free-fall (large statistical samples).
Over the last few years these attributes have largely been realized, with optical lattice clocks being demonstrated at $1\times10^{-16}$ fractional uncertainty~\cite{LudZelCam08etal}. On a theoretical basis it is anticipated that these clocks will further reach to the $10^{-18}$ level \cite{HacMiyPor08etal,KatHasIli09}. To achieve such an unprecedented accuracy 
it is imperative that all frequency shifts induced by the lattice laser field are well understood and evaluated at this level.
A number of important theoretical works have been dedicated to determining or estimating frequency shifts arising from mechanisms such as hyperpolarizability (second order in intensity) \cite{KatTakPal03,TaiYudOvs06}, black-body radiation \cite{PorDer06}, collisions \cite{Gib09}, off-resonant couplings to the probe laser \cite{TaiYudOat10etal}, and magnetic dipole and electric quadrupole interactions with the lattice field \cite{TaiYudOvs08,KatHasIli09}.

All optical lattice clocks which have been demonstrated to this point operate at lattice wavelengths which are detuned to the red of a nearby atomic resonance.
Proposed clocks based on blue-detuned lattices are presumed to be inherently less susceptible to lattice-induced frequency shifts~\cite{TakKatMar09}. This assumption is based on the fact that atoms in a blue-detuned lattice are localized to regions of intensity minima rather than intensity maxima, thus suppressing higher order hyperpolarizability effects and scattering of lattice photons. Takamoto \emph{et al.}~\cite{TakKatMar09} discussed the prospects of a Sr clock with a blue-detuned lattice and have experimentally identified an optimal magic wavelength at 390 nm.
Suitable blue magic wavelengths are also likely to exist in other alkaline-earth(-like) species (e.g., Mg, Ca, Yb).
To exemplify the common conviction that blue-detuned optical lattice clocks will ultimately provide superior accuracy to their red-detuned counterparts, we refer to recent Letter~\cite{KatHasIli09} in which it is argued that clock uncertainty at the $10^{-18}$ level will be within reach; we note that this conjecture is preceded by the caveat of operation at a blue-detuned magic wavelength.

In their original proposal, Katori~\emph{et al.}~suggested the possibility of using 3-D lattices with single atom occupancy (or less) at each site, thus effectively eliminating collisional effects. Recently a 3-D lattice clock has been realized with bosonic $^{88}$Sr atoms~\cite{AkaTakKat10}, marking another important step towards achieving the anticipated accuracy of the lattice clocks.

Here we consider the frequency shift in optical lattice clocks arising from the coupling of the electronic motion to the atomic motion within the lattice; we refer to this as the non-adiabatic shift. For the simplest 3-D lattice geometry we find that this coupling only renders a frequency shift in the blue-detuned lattice clocks. We estimate the size of the non-adiabatic shift for the prospective ``blue'' Sr lattice clock and find it to be on the order of $10^{-18}$ fractionally. In this particular case, the non-adiabatic shift is largely overshadowed by the electric quadrupole shift, though for other geometries or for clocks based on other atomic species it could perhaps provide the leading source of clock uncertainty.

%


We begin our analysis with the total Hamiltonian for an atomic system in the presence of some external laser configuration (e.g., an optical lattice),
\begin{equation*}
H(\mathbf{r},t,\overline{q})=T(\mathbf{r})+H_{\mathrm{at}}(\overline{q})+H_\mathrm{L}(\mathbf{r},t,\overline{q}),
\end{equation*}
where $\mathbf{r}$ denotes the atomic center of mass and $\overline{q}$ encapsulates all electronic spatial coordinates relative to the center of mass as well as 
spin coordinates. Here $T(\mathbf{r})=-(\hbar^2/2M)\nabla^2$ represents the kinetic energy associated with the center of mass motion ($M$ being the mass of the atom), $H_{\mathrm{at}}(\overline{q})$ is the conventional field-free atomic Hamiltonian
(including electronic kinetic energy, Coulomb interactions, Breit interactions, etc.),
and $H_\mathrm{L}(\mathbf{r},t,\overline{q})$ represents the interaction of the atom with the laser field. We treat the laser field classically, and as such the operator $H_\mathrm{L}(\mathbf{r},t,\overline{q})$ is a simple function of $\mathbf{r}$ and $t$ (i.e., it does not contain derivatives with respect to these coordinates); furthermore it is assumed to be periodic in time with period $\tau$,
$
H_\mathrm{L}(\mathbf{r},t+\tau,\overline{q})=H_\mathrm{L}(\mathbf{r},t,\overline{q}).
$
It follows that the Hamiltonian $H(\mathbf{r},t,\overline{q})$ is also periodic in time with period $\tau$.

Due to the periodicity of the Hamiltonian, we may treat this problem within the framework of the well-known Floquet formalism~\cite{Sam73}, in which the evolution of the system is governed by the eigenvalue equation
\begin{equation}
\mathcal{H}(\mathbf{r},t,\overline{q})\psi(\mathbf{r},t,\overline{q})=\mathcal{E}\psi(\mathbf{r},t,\overline{q}).
\label{Eq:SchroRecast}
\end{equation}
where $\mathcal{H}(\mathbf{r},t,\overline{q})\equiv H(\mathbf{r},t,\overline{q})-i\hbar\frac{\partial}{\partial t}$ and $\psi(\mathbf{r},t,\overline{q})$ is periodic in time with period $\tau$. With the total wave function given by $\Psi(\mathbf{r},t,\overline{q})=\psi(\mathbf{r},t,\overline{q})e^{-i\mathcal{E}t/\hbar}$, the eigenvalue equation (\ref{Eq:SchroRecast}) is entirely equivalent to the time-dependent Schr{\"o}dinger equation.
The eigenvalue $\mathcal{E}$ corresponds to the time-averaged energy of the state $\Psi(\mathbf{r},t,\overline{q})$ \cite{LanEpsKar72} and, for an optical lattice clock, therefore represents the physically significant energy level being probed.
We will refer to $\mathcal{H}(\mathbf{r},t,\overline{q})$ as the steady state Hamiltonian and Eq.~(\ref{Eq:SchroRecast}) as the steady state Schr{\"o}dinger equation.

We wish to find approximate solutions to Eq.~(\ref{Eq:SchroRecast}). To this end, we begin by considering the Hamiltonian exclusive of the kinetic energy term $T(\mathbf{r})$. In particular, we look for solutions to the eigenvalue equation
\begin{equation}
\mathcal{H}_\mathrm{at,L}(\mathbf{r},t,\overline{q})\chi_n(\mathbf{r},t,\overline{q})=\mathcal{W}_n(\mathbf{r})\chi_n(\mathbf{r},t,\overline{q}),
\label{Eq:atLssHam}
\end{equation}
where the atom-laser steady state Hamiltonian is
$\mathcal{H}_\mathrm{at,L}(\mathbf{r},t,\overline{q})\equiv\mathcal{H}(\mathbf{r},t,\overline{q})-T(\mathbf{r})$
and we have introduced the quantum number $n$ to label the eigensolutions.
Effectively the atom-laser steady state Hamiltonian describes an atom and its interaction with the laser field when the atom is at a \emph{fixed} location $\mathbf{r}$ within the laser field.
Below we will consider solutions to Eq.~(\ref{Eq:atLssHam}) in the context of steady state perturbation theory; for the time being we will assume the eigenvalues and eigenfunctions are known. We will, however, find it useful at this point to separate the eigenvalues into two terms,
\begin{equation}
\mathcal{W}_n(\mathbf{r})=W_n+U_n(\mathbf{r}),
\label{Eq:eigensplit}
\end{equation}
where $W_n$ is the energy of the atom in the field-free limit ($H_\mathrm{L}(\mathbf{r},t,\overline{q})\rightarrow0$) and $U_n(\mathbf{r})$ is then the corresponding level shift induced by the presence of the laser field.

As the second piece of our approximation we consider the motion of the atom within the laser field as governed by the time-independent eigenvalue equation
\begin{equation*}
\left[T(\mathbf{r})+U_n(\mathbf{r})\right]\varphi_{nv}(\mathbf{r})=G_{nv}\varphi_{nv}(\mathbf{r}),
\end{equation*}
with the eigensolutions (for a specific atomic state $n$) being labeled by the quantum number $v$. We assume these eigensolutions to be known as well.

We now graft our decoupled eigenfunctions together,
\begin{equation*}
\psi_{nv}(\mathbf{r},t,\overline{q})=\chi_n(\mathbf{r},t,\overline{q})\varphi_{nv}(\mathbf{r}).
\end{equation*}
With the approximation
$
\nabla^2\chi_n(\mathbf{r},t,\overline{q})\varphi_{nv}(\mathbf{r})\cong\chi_n(\mathbf{r},t,\overline{q})\nabla^2\varphi_{nv}(\mathbf{r}),
$
it subsequently follows that
\begin{equation*}
\mathcal{H}(\mathbf{r},t,\overline{q})\psi_{nv}(\mathbf{r},t,\overline{q})\cong\left(W_n+G_{nv}\right)\psi_{nv}(\mathbf{r},t,\overline{q}),
\end{equation*}
i.e., the $\psi_{nv}(\mathbf{r},t,\overline{q})$ represent approximate eigenfunctions of the steady state Hamiltonian, Eq.~(\ref{Eq:SchroRecast}), with corresponding eigenvalues $\left(W_n+G_{nv}\right)$. Generally speaking, for atoms trapped in an optical lattice this approximation is quite good and can be justified by the fact that the electron mass is much smaller than the nuclear mass.
In physical terms, as an atom moves within a lattice site it experiences a varying lattice field and the electronic response to this varying field occurs on a time scale much faster than the atomic motion (the atomic motion effectively being adiabatic).
We will refer to the correction which accounts for coupling of the electronic motion to atomic motion the non-adiabatic correction. We emphasize that the non-adiabatic effects discussed here are not associated with any type of dynamical change in the optical lattice field (e.g., ramping up or down the intensity).

We aim to estimate the size of the non-adiabatic correction. Analogous to the time-independent Schr{\"o}dinger equation, the steady state Schr{\"o}dinger equation is derivable from a variational formalism \cite{Sam73}, and as such we may estimate the true value of $\mathcal{E}$ with an appropriate trial function, namely the approximate (``adiabatic'') eigenfunctions $\psi_{nv}(\mathbf{r},t,\overline{q})$. A reasonable estimate of the non-adiabatic correction, $F_{nv}$, is then given by (we assume the adiabatic eigenfunctions to be normalized)
\begin{equation}
F_{nv}
=\langle\langle\langle\psi_{nv}|\mathcal{H}|\psi_{nv}\rangle\rangle\rangle-\left(W_n+G_{nv}\right),
\label{Eq:DE}
\end{equation}
where the somewhat cumbersome yet useful bra-ket notation with triple-angled brackets indicates integration over $\mathbf{r}$, $t$, and $\overline{q}$, with the $t$-integration over one period,
\begin{eqnarray*}
\langle\langle\langle\psi|\mathcal{H}|\psi\rangle\rangle\rangle
&\equiv&\frac{1}{\tau}\int\int_{-\tau/2}^{+\tau/2}\int d^3\mathbf{r}dtd\overline{q}
\\&&\times
\psi^*(\mathbf{r},t,\overline{q})\mathcal{H}(\mathbf{r},t,\overline{q})\psi(\mathbf{r},t,\overline{q}).
\end{eqnarray*}
We will also find it useful to designate the double-angled brackets $\langle\langle\dots\rangle\rangle$ for integrations over $t$ and $\overline{q}$ and single-angled brackets $\langle\dots\rangle$ for integrations over $\overline{q}$ alone.
Eq.~(\ref{Eq:DE}) reduces to the form
\begin{eqnarray}
F_{nv}
&=&-\frac{\hbar^2}{2M}\int d^3\mathbf{r}\left[\varphi_{nv}^*(\mathbf{r})\varphi_{nv}(\mathbf{r})\langle\langle\chi_n|\nabla^2|\chi_n\rangle\rangle
(\mathbf{r})\right.
\nonumber\\
&&\left.
+2\varphi_{nv}^*(\mathbf{r})\mathbf{\nabla}\varphi_{nv}(\mathbf{r})\cdot
\langle\langle\chi_n|\mathbf{\nabla}|\chi_n\rangle\rangle(\mathbf{r})\right],
\label{Eq:Ecorr}
\end{eqnarray}
where the notation $\langle\langle\chi_n|\nabla^2|\chi_n\rangle\rangle
(\mathbf{r})$ and $\langle\langle\chi_n|\mathbf{\nabla}|\chi_n\rangle\rangle(\mathbf{r})$ emphasizes the fact that these are functions of $\mathbf{r}$.

Ultimately we are interested in the non-adiabatic shift of the clock frequency. The clock probes the energy intervals
\begin{equation*}
\Delta\mathcal{E}_{v}
=\Delta W+ \Delta G_{v} + \Delta F_{v},
\end{equation*}
where $\Delta$, here and below, is used to denote the difference between clock levels, e.g., $\Delta\mathcal{E}_{v}\equiv \mathcal{E}_{P,v}-\mathcal{E}_{S,v}$ for clock states ${^3P_0}$ and ${^1S_0}$, and we assume the usual case of $v\rightarrow v$ transitions. Here $\Delta W$ is the unperturbed interval and the combination $\Delta G_{v} + \Delta F_{v}$ gives the lattice-induced perturbation leading to a shift in the measured clock frequency.


To estimate the size of the shift $\Delta F_{v}$, we return to the atom-laser steady state Schr{\"o}dinger equation, Eq.~(\ref{Eq:atLssHam}).
Eigensolutions of Eq.~(\ref{Eq:atLssHam}) can be obtained by a systematic application of steady state (Floquet) perturbation theory \cite{Sam73}, treating the interaction $H_\mathrm{L}(\mathbf{r},t,\overline{q})$ as a perturbation. The zeroth order atom-laser steady state Hamiltonian,
$\mathcal{H}^{(0)}_\mathrm{at,L}(t,\overline{q})
\equiv\mathcal{H}_\mathrm{at,L}(\mathbf{r},t,\overline{q})-H_\mathrm{L}(\mathbf{r},t,\overline{q})$,
has eigenvalues and eigenfunctions satisfying
\begin{equation}
\mathcal{H}^{(0)}_\mathrm{at,L}(t,\overline{q})\chi_n^{(0)}(\overline{q})=\mathcal{W}_n^{(0)}\chi_n^{(0)}(\overline{q}).
\label{Eq:atomlaserSchroZero}
\end{equation}
We note the absence of $\mathbf{r}$ in this last expression. The zeroth order eigenvalues and eigenfunctions are given by
\begin{equation*}
\mathcal{W}^{(0)}_{n}=W_n,\qquad \chi^{(0)}_{n}(\overline{q})=f_{n}(\overline{q}),
\end{equation*}
where $W_n$ and $f_n(\overline{q})$ are the eigenvalues and eigenfunctions of the time-independent Schr{\"o}dinger equation for the field-free atom,
$
H_{\mathrm{at}}(\overline{q})f_n(\overline{q})=W_nf_n(\overline{q}).
$

A complete set of eigensolutions to Eq.~(\ref{Eq:atomlaserSchroZero}) (spanning the space of all functions square integrable over $\overline{q}$ and periodic in time with period $\tau$) is provided by
\begin{equation}
\mathcal{W}^{(0)}_{np}=W_n+p\hbar\omega,\qquad \chi^{(0)}_{np}(t,\overline{q})=f_{n}(\overline{q})e^{ip\omega t},
\label{Eq:HilbertComplete}
\end{equation}
where $\omega=2\pi/\tau$ and the additional quantum number $p$ takes the value of any integer. The various values of $p$ do not correspond to physically distinct states (i.e., the wave function $\chi^{(0)}_{np}(t,\overline{q})e^{-i\mathcal{W}^{(0)}_{np}t/\hbar}$ is independent of $p$), and we obtain a one-to-one correspondence with atomic states by setting $p=0$ above. However, in the application of steady state perturbation theory we require summations over a complete set of states, and this is furnished by Eqs.~(\ref{Eq:HilbertComplete}) with all integer $p$. We note that these are the familiar ``dressed'' atomic states (in which case $p$ is simply associated with photon number).

At this point it is useful to consider a specific form for the interaction $H_\mathrm{L}(\mathbf{r},t,\overline{q})$. Initially we will limit our attention to the dominant electric dipole ($E1$) interaction.
The $E1$ interaction effectively ``samples'' the electric field at the location of the atom and has the form
\begin{equation*}
H_\mathrm{L}(\mathbf{r},t,\overline{q})=-\mathrm{Re}\left[\mathbf{E}(\mathbf{r})e^{-i\omega t}\right]\cdot\mathbf{D}(\overline{q}),
\end{equation*}
where $\mathbf{D}(\overline{q})$ is the atomic dipole operator and $\mathrm{Re}\left[\mathbf{E}(\mathbf{r})e^{-i\omega t}\right]$ is the (classical) electric field with an assumed harmonic time dependence and complex amplitude $\mathbf{E}(\mathbf{r})$.

The application of steady state perturbation theory proceeds in the same manner as conventional time-independent perturbation theory. The first order correction to the eigenvalue, $\mathcal{W}^{(1)}_{n}(\mathbf{r})=\langle\langle \chi^{(0)}_{n}|H_\mathrm{L}|\chi^{(0)}_{n}\rangle\rangle$, vanishes when the integration over time is performed (and, within the $E1$ approximation, from parity considerations as well). After integrating over time, the second order correction reads
\begin{equation}
\mathcal{W}^{(2)}_{n}(\mathbf{r})=\frac{1}{4}\sum_{n^\prime}\left\{\frac{\left[\mathbf{E}^*(\mathbf{r})\cdot\langle n|\mathbf{D}|n^\prime \rangle\right]\left[\mathbf{E}(\mathbf{r})\cdot\langle n^\prime|\mathbf{D}|n\rangle\right]}{W_{n}-W_{n^\prime}+\hbar\omega}\right\}^\circledast
\!\!\!,
\label{Eq:2ndorder}
\end{equation}
where $|n\rangle\equiv|f_n\rangle$ and the notation $\{\dots\}^\circledast$ indicates that one should add to the term in braces a complementary term acquired by making the simultaneous substitutions $\mathbf{E}(\mathbf{r})\rightarrow\mathbf{E}^*(\mathbf{r})$ and $\omega\rightarrow-\omega$. We will not consider higher order (hyperpolarizability) corrections, and so in line with Eq.~(\ref{Eq:eigensplit}), we set $U_n(\mathbf{r})=\mathcal{W}^{(2)}_{n}(\mathbf{r})$.

For scalar atomic states (e.g., clock states ${^1S_0}$, ${^3P_0}$) Eq.~(\ref{Eq:2ndorder}) may be recoupled \cite{VarMosKhe88} and expressed as
\begin{equation}
U_n(\mathbf{r})=-\bar{\alpha}_nI(\mathbf{r}),
\label{Eq:U}
\end{equation}
where $I(\mathbf{r})=(c/8\pi)\left|\mathbf{E}(\mathbf{r})\right|^2$ is the intensity at $\mathbf{r}$ and $\alpha_n=(c/2\pi)\bar{\alpha}_n$ is the conventional (scalar) ac polorizability,
\begin{equation}
\alpha_n=-\frac{1}{3}\sum_{n^\prime}\left|\langle n^\prime|\mathbf{D}|n\rangle\right|^2
\left\{\frac{1}{W_{n}-W_{n^\prime}+\hbar\omega}\right\}^\circledast,
\label{Eq:alpha}
\end{equation}
with $c$ being the speed of light.
We note the $\omega$-dependence of the ac polarizability.


After integration over time, the first order correction for the eigenfunction $\chi_{n}(\mathbf{r},t,\overline{q})$ is found to be
\begin{equation*}
\chi^{(1)}_{n}(\mathbf{r},t,\overline{q})=
-\frac{1}{2}\sum_{n^\prime}\left\{\frac{\left[\mathbf{E}(\mathbf{r})\cdot\langle n^\prime|\mathbf{D}|n\rangle\right]}{W_{n}-W_{n^\prime}+\hbar\omega}e^{-i\omega t}
\right\}^\circledast f_{n^\prime}(\overline{q}).
\end{equation*}
For some operator $S(\mathbf{r})$, the leading correction to $\langle\langle\chi_{n}|S|\chi_{n}\rangle\rangle(\mathbf{r})$ is given by $\langle\langle\chi^{(1)}_{n}|S|\chi^{(1)}_{n}\rangle\rangle(\mathbf{r})$. In particular, we are interested in the cases
$S(\mathbf{r})=\nabla^2$ and $S(\mathbf{r})=\mathbf{\nabla}$.
We now assume that the lattice is composed of a superposition of standing waves of linear polarization, with all standing waves in phase (e.g., the ``folded'' 3-D optical lattice in Ref.~\cite{AkaTakKat10}). In such a lattice arrangement $\mathbf{E}(\mathbf{r})$ may be taken as real and we find the following expressions for scalar atomic states after recoupling,
\begin{eqnarray}
\langle\langle\chi^{(1)}_{n}|\nabla^2|\chi^{(1)}_{n}\rangle\rangle(\mathbf{r})&=&-\left(\frac{\omega}{c}\right)^2
\bar{\sigma}_nI(\mathbf{r}),
\nonumber\\
\langle\langle\chi^{(1)}_{n}|\mathbf{\nabla}|\chi^{(1)}_{n}\rangle\rangle(\mathbf{r})&=&
\frac{1}{2}\bar{\sigma}_n\mathbf{\nabla}I(\mathbf{r}),
\label{Eq:chicorrections}
\end{eqnarray}
where the frequency-dependent atomic factor $\sigma_n=(c/2\pi)\bar{\sigma}_n$ is given by
\begin{equation}
\sigma_n=\frac{1}{3}\sum_{n^\prime}\left|\langle n^\prime|\mathbf{D}|n\rangle\right|^2
\left\{\frac{1}{\left(W_{n}-W_{n^\prime}+\hbar\omega\right)^2}\right\}^\circledast.
\label{Eq:sigma}
\end{equation}

We now assume a specific lattice geometry; we choose the most straightforward 3-D geometry (case (I) of Ref.~\cite{KatHasIli09}) in which the lattice is composed of three standing waves aligned in the $x$, $y$, and $z$ directions having mutually orthogonal linear polarizations. It follows that the intensity is given by
\begin{equation}
I(\mathbf{r})=
\begin{cases}
4\left[I_x\mathrm{cos}^2\left(\frac{\omega}{c}x\right)+I_y\mathrm{cos}^2\left(\frac{\omega}{c}y\right)+I_z\mathrm{cos}^2\left(\frac{\omega}{c}z\right)\right]
\\
4\left[I_x\mathrm{sin}^2\left(\frac{\omega}{c}x\right)+I_y\mathrm{sin}^2\left(\frac{\omega}{c}y\right)+I_z\mathrm{sin}^2\left(\frac{\omega}{c}z\right)\right]
\end{cases}
\!\!\!\!,
\label{Eq:I}
\end{equation}
where have chosen the coordinate origin separately for the cases of red (top) and blue (bottom) lattices such that $U_n(\mathbf{r})$ (Eq.~(\ref{Eq:U})) has a minimum at the origin.

We now Taylor expand Eq.~(\ref{Eq:I}) to quadratic terms in $x$, $y$, and $z$ and evaluate the non-adiabatic correction (Eq.~(\ref{Eq:Ecorr})) using Eqs.~(\ref{Eq:chicorrections}) and assuming harmonic oscillator vibrational functions
$\varphi_{nv}(\mathbf{r})$ with $v=(v_x,v_y,v_z)$; the resulting expression reads
\begin{equation}
F_{nv}
=\ell12R\bar{\sigma}_n I_0
+(-1)^{\ell+1}\frac{2R^{3/2}\bar{\sigma}_n}{\sqrt{\left|\bar{\alpha}_n\right|}}\sum_i\sqrt{I_i}\left(v_i+{\textstyle \frac{1}{2}}\right),
\label{Eq:Fnv}
\end{equation}
where $I_0\equiv (I_x+I_y+I_z)/3$, $R\equiv\hbar^2(\omega/c)^2/2M$ is the recoil energy, and $\ell$ is equated to 0 and 1 for red and blue lattices, respectively.

To understand the influence of $F_{nv}$ 
on the clock frequency, it pays to consider an explicit form for the vibrational energies $G_{nv}$ as well.
Taichenachev~\emph{et al.}~\cite{TaiYudOvs08} have shown that magnetic dipole ($M1$) and electric quadrupole ($E2$) interactions can give non-negligible contributions to $G_{nv}$, and so we include these effects in our consideration. For the present lattice arrangement, $G_{nv}$ is given by~\cite{KatHasIli09}
\begin{equation}
G_{nv}
=-12\bar{\alpha}_n^\prime I_0
+4{\sqrt{R\left|\bar{\alpha}_n\right|}}\sum_i\sqrt{I_i}\left(v_i+{\textstyle \frac{1}{2}}\right),
\label{Eq:Gnv}
\end{equation}
where again we have employed the harmonic oscillator approximation. Here $\bar{\alpha}_n$ now includes $M1$ and $E2$ effects in addition to the dominant $E1$ effects ($\bar{\alpha}_n=\bar{\alpha}_n^{E1}-\bar{\alpha}_n^{M1}-\bar{\alpha}_n^{E2}$) and $\bar{\alpha}_n^\prime$ is the small corrective difference ($\bar{\alpha}_n^\prime=\bar{\alpha}_n^{M1}+\bar{\alpha}_n^{E2}$); this slightly modified $\bar{\alpha}_n$ is appropriate in Eq.~(\ref{Eq:Fnv}) as well.
We note the simlarities between $G_{nv}$ and $F_{nv}$: 
both expressions consist of two terms, with the first term being independent of vibrational state and proportional to $I_0$, while the second term is proportional to $\sum_i\sqrt{I_i}\left(v_i+{\textstyle \frac{1}{2}}\right)$.

The relative shift $\Delta{G}_v+\Delta{F}_v$ is dependent upon $\omega$ via the atomic factors $\bar{\alpha}_n$, $\bar{\alpha}_n^\prime$, and $\bar{\sigma}_n$; thus the lattice induced frequency shift may be modulated by tuning the lattice lasers to an appropriate wavelength, the so-called magic wavelength. With the intent of eliminating the $M1$ and $E2$ effects in lattice clocks which scale with the square-root of intensity (second term of Eq.~(\ref{Eq:Gnv})), Katori~\emph{et al.}~\cite{KatHasIli09} recently defined the magic wavelength by the criteria $\Delta\bar{\alpha}=0$ and proposed experimental determination of it by analyzing, and minimizing to zero, the $v$-dependence of $\Delta{\mathcal{E}}_v$. We see here that such an experimental procedure actually absorbs the second term of Eq.~(\ref{Eq:Fnv}) as well;
the definition of the magic wavelength is naturally extended to eliminate the net (vibrational and non-adiabatic) effects which scale with the square-root of intensity.
For operation at this magic wavelength, the remaining terms which give rise to a frequency shift are the first terms of Eqs.~(\ref{Eq:Fnv},\ref{Eq:Gnv}),
\begin{equation*}
\Delta G + \Delta F
=12I_0\left(-\Delta\bar{\alpha}^\prime +\ell R\Delta\bar{\sigma}\right).
\end{equation*}
An interesting result is that the non-adiabatic shift only affects blue-detuned ($\ell=1$) lattice clocks to within our approximations.

We now estimate the size of the shifts $\Delta G$ and $\Delta F$ for the prospective Sr lattice clock operating at the 390 nm blue magic wavelength. We find a total shift
\begin{equation}
\Delta G + \Delta F
=12|\bar{\alpha}|I_0\left(1.4\times10^{-7}+9.5\times10^{-10}\right),
\end{equation}
where $\bar{\alpha}_P\cong\bar{\alpha}_S\equiv\bar{\alpha}$ and the two terms on the right-hand-side are respective contributions. To obtain this result we have borrowed the ratio $\Delta\bar{\alpha}^\prime/|\bar{\alpha}|=-1.4\times10^{-7}$ from Ref.~\cite{KatHasIli09} and we have evaluated equations (\ref{Eq:alpha},\ref{Eq:sigma}) using dipole matrix elements from Ref.~\cite{ZhoXuChe10} and energies from Ref.~\cite{Moo71} ($\alpha_S=\alpha_P=-455~\mathrm{a.u.}$, $\sigma_S=28000~\mathrm{a.u.}$, $\sigma_P=218000~\mathrm{a.u.}$). Assuming $I_x=I_y=I_z=I_0$ with a potential depth of $10~\mu\mathrm{K}$ (the corresponding potential depth-to-recoil energy ratio being $4|\bar{\alpha}|I_0/R=14$), we find the following fractional shifts to the clock frequency
\begin{equation*}
\frac{\delta\nu}{\nu}=\frac{\Delta G + \Delta F}{\Delta W}
=2.0\times10^{-16}+1.4\times10^{-18}.
\end{equation*}
We see that the non-adiabatic shift is on the order of $10^{-18}$, accounting for about 0.7\% of the total shift. This is on the same order as the $M1$ contribution (0.8\%~\cite{KatHasIli09}), both of these effects being overshadowed by the $E2$ contribution.

We note that for the assumed trap depth and lattice geometry, clock accuracy at the $10^{-18}$ level can only be achieved if the combined parameter $\Delta\bar{\alpha}^\prime I_0$ is known/controlled to better than 5\%.
This 
may prove difficult to attain experimentally. In Ref.~\cite{KatHasIli09}, two novel lattice geometries were conceived such as to effectively eliminate either the $E2$ (case (II)) or the $M1$ (case (III)) frequency shift. For the blue Sr lattice clock, implementation of the case (II) geometry may be necessary to realize accuracy at the $10^{-18}$ level. In such a set-up, the non-adiabatic and $M1$ contributions are no longer masked by the $E2$ shift and are likely to constitute the leading uncertainty in the clock.

It is also worth noting that the red-detuned 3-D lattice clock of Ref.~\cite{AkaTakKat10} does not have the specific geometry considered above and thus may be susceptible to the non-adiabatic effects.

If we treat the potential depth-to-recoil energy ratio as a fixed parameter, then the non-adiabatic shift is found to scale as $R^2$. Thus we reason that the non-adiabatic effects may be enhanced for blue lattice clocks based on other atomic species. Considering the lattice clock candidate Mg (see, e.g., Ref.~\cite{FriPapRie08etal}), we note that the atomic mass and energy spectrum necessitate a larger recoil energy than the blue Sr lattice clock ($R_\mathrm{Mg}\gtrsim7R_\mathrm{Sr}$) and the resulting non-adiabatic shift could easily reach to the $10^{-16}$ level.
We further note that the non-adiabatic effects could potentially be enhanced for blue magic wavelengths which are relatively close to atomic resonance, as is evident from the approximate relation $\sigma_n/\alpha_n\approx-1/\delta_n$, with $\delta_n$ being the energy detuning from resonance.




%

This work was supported by the Marsden Fund 
(Royal Society of New Zealand). The author thanks A. Derevianko and P. Schwerdtfeger for useful comments.


%

\end{document}